# Hamiltonian Cosmology

**M Ibison**

Institute for Advanced Studies at Austin,

11855 Research Boulevard, Austin TX 78759-2443, USA

E-mail: ibison@earthtech.org

**Abstract.** The cosmological scale factor $a(t)$ of the flat-space Robertson-Walker geometry is examined from a Hamiltonian perspective wherein $a(t)$ is interpreted as an independent dynamical coordinate and the curvature density $\sqrt{-g(a)}R(a,\dot{a},\ddot{a})$ is regarded as an action density in Minkowski spacetime. The resulting Hamiltonian for $a(t)$ is just the first Friedmann equation of the traditional approach (i.e. the Robertson-Walker cosmology of General Relativity) – as might be expected. The utility of this approach however derives from the fact that each of the terms - matter, radiation, and vacuum, including the kinetic / gravitational field term – are formally energy densities, and the equation as a whole becomes a formal statement of energy conservation. An advantage of this approach is that it facilitates an intuitive understanding of energy balance and exchange on the cosmological scale that is otherwise absent in the traditional presentation. Each coordinate system has its own – internally consistent - explanation for how energy balance is achieved. For example, in the spacetime with line element $ds^2 = dt^2 - a^2(t)d\mathbf{x}^2$, cosmological red-shift emerges as due to a post-recombination interaction between the scalar field $a(t)$ and the EM fields in which the latter loose energy as if propagating through a homogeneous lossy medium, with the energy lost to the scale factor helping drive the cosmological expansion.

**1. Introduction**

Since Cosmological metrics are time-dependent there is no time-like Killing vector, and so no corresponding means of identifying a conserved energy. In their books on cosmology, Harrison [1] and Peebles [2] goes as far as stating that energy is simply not conserved in general relativity. Baryshev [3] recently drew attention to the role of cosmological red-shift in frustrating attempts to give meaning to energy conservation in cosmology.

Harrison has presented a comprehensive Newtonian picture of cosmology consistent with the flat space Robertson-Walker spacetime Eq. (1). The resulting Friedmann equation (see Eq. (25) below) is interpreted as the first integral of Newton's second law, wherein the Hubble term proportional to $\dot{a}^2$ is interpreted as a kinetic energy. An advantage of this picture of the Friedmann equation is that it complies with the ordinary flat-space intuition of energy conservation. A disadvantage is that there is no formal basis in GR for granting the Hubble term the status of an energy density - on equal footing with matter, vacuum and other fields. The following can be regarded as a development of the Newton-Harrison picture to put it on a solid mathematical foundation by providing a formal connection with GR. In that process we are able identify the Hubble term in the Friedmann equation as the energy density of the cosmological gravitational field - when appropriately scaled according to the coordinate system.

Due to the general coordinate-invariance of GR one does not expect to be able to assign a coordinate-independent interpretation to the various contributing terms (matter, radiation, vacuum) in the Friedmann equation. Even so, the Hamiltonian formulation is possible because once a coordinate choice has been made, one may then speak formally and unambiguously of energy conservation in that coordinate system, so that the various terms - including the gravitational field - can be interpreted as energy densities in a self-consistent manner. Also, after exploring the various alternative coordinate-dependent interpretations, in the end the method promotes a deeper coordinate-independent understanding of role and *relative* status of the contributing terms.

Here we will work almost exclusively with the flat-space Robertson-Walker (RW) spacetime. We will refer to the coordinate system in which the line-element is

$$ds^2 = dt_{RW}^2 - a^2(t_{RW})d\mathbf{x}^2 \tag{1}$$

as the 'traditional' (RW) representation of that spacetime. The principal idea here is to regard the scale factor $a(t)$ (in any representation) as a dynamical variable in Minkowski spacetime, and then call upon the Hamilton formalism to extract the Hamiltonian for the whole system directly, including the 'gravitational field', and without consideration of the Euler equations. Lest there be some resistance to the mention of a Minkowski background, we draw attention to the fact that the variational procedure by which the Einstein equation is derived from the Hilbert action is insensitive to the interpretation placed upon the metric components as either potentials in Minkowski spacetime [1] or descriptors of the geometry of curved spacetime - a point emphasized by Feynman [4]. This remains the case even for the curved-space $K \neq 0$ RW spacetimes, despite the fact that the (global) topology of space at fixed universal time is not the flat Euclidean space of Minkowski spacetime. Even so, here we restrict attention to the flat space RW spacetime, avoiding concerns over incompatible topologies.

We go over to a Minkowski picture by interpreting the metric tensor as a *Lorentz tensor field* whose components are dynamical variables in the $R(1,3)$ Minkowski space spanned by the particular coordinates whose line element is $ds^2 = dt^2 - d\mathbf{x}^2$. The corresponding canonical Hamiltonian will be the total energy density of the system whose integral over all space is an (arbitrary) constant of the motion. The definition is not of course unique - it depends on the coordinate system. The possibilities considered here are: the standard (RW) coordinate system, the conformal coordinate system, a coordinate system which reproduces the Newtonian Cosmology, and a coordinate system with constant 4-volume (unit determinant). In section 8 this method is placed in the more general context of techniques for extracting a gravitational stress-energy tensor.

## 2. Hamiltonian in constant volume coordinates

*2.1. Cosmological action*

Though the total action with metric Eq. (1) for a cosmological fluid will be very well known, here we take the trouble to write down the contributions in a coordinate independent manner. Post recombination, the matter and EM fields are decoupled, so the action is just the sum of gravitational, vacuum, matter and EM contributions:

$$L = L_g + L_v + L_m + L_{EM} \,. \tag{2}$$

Einstein GR is invoked by choosing the Hilbert action

$$L_g = \frac{1}{16\pi G} \int d^4x \sqrt{-g} R \tag{3}$$

where $R$ is the Ricci scalar curvature. The vacuum is just

---

[1] Strictly, without a suitable redefinition, the $g_{ab}(x)$ would be potentials in a 4D spacetime with *Euclidean* signature. For a diagonal metric such as considered here, one may prefer to work with potentials $\phi_{ab}(x)$ such that $g_{ab}(x) = \eta_{ab}\phi_{ab}(x)$ (no sum) where $\eta_{ab}$ is the Lorentz-Minkowski metric. But see also section 8.

$$L_v = -\rho_{v0} \int d^4x \sqrt{-g} \tag{4}$$

where $\rho_{v0}$ is the constant coordinate density. In general, the classical matter action is

$$L_m = -\sum_i m_i \int ds \sqrt{g_{ab} \frac{dx_i^a}{ds} \frac{dx_i^b}{ds}} . \tag{5}$$

Here we consider particles that are static in the co-moving frame, whereupon

$$L_m = -\sum_i m_i \int ds \frac{dx_{(i)}^0}{ds} \sqrt{g_{00}} = -\sum_i m_{(i)} \int dx_{(i)}^0 \sqrt{g_{00}(x_{(i)})} \tag{6}$$

where the sum is over a fixed number of particles in a fixed proper volume. In the continuum limit, and assuming a uniform distribution,

$$\sum_i m_{(i)} \to \rho_{m0} \int d^3\mathbf{x}_{(i)} \tag{7}$$

where $\rho_{m0}$ is the (coordinate) density of matter at some initial time. With this Eq. (6) can be written

$$L_m = -\rho_{m0} \int d^4x \sqrt{g_{00}(x)} . \tag{8}$$

The effect of the cosmological gravitational field on $1^{st}$ quantized – rather than classical – matter is considered briefly in Appendix A. The EM action in curved spacetime is

$$L_{EM} = -\frac{1}{4} \int d^4x \sqrt{-g} F_{ab} F^{ab} = -\frac{1}{4} \int d^4x \sqrt{-g} F_{ab} F_{cd} g^{ac} g^{bd} \tag{9}$$

where the $\{F_{ab}\}$ (and $\{g_{ab}\}$) are generally relativistic tensors. Combining Eqs. (3), (4), (8), and (9), the total action corresponding to the $K=0$ RW Cosmology is

$$L = -\int d^4x \left\{ -\frac{1}{16\pi G} \sqrt{-g} R + \sqrt{-g} \rho_{v0} + \rho_{m0} \sqrt{g_{00}} + \frac{1}{4} \sqrt{-g} F_{ab} F_{cd} g^{ac} g^{bd} \right\} . \tag{10}$$

*2.2. Effective action in Minkowski spacetime*

We choose now to present the analysis in detail for the constant volume coordinate system defined by $\sqrt{-g} = 1$. The results for the other systems are easily inferred and where required the results will be written down directly. Analysis in the constant-volume system has the advantage of providing some contrast and perspective to the more familiar RW and conformal systems, particularly with regard to the evolution of the EM energy density.

The constant volume system is achieved by the transformation of the RW system Eq. (1) having time $t_{RW}$ to a new time coordinate $t_{cv}$:

$$t_{cv} = \int dt_{RW} a^3 \Rightarrow \{g_{\mu\nu}\} = diag(1/a^6, -a^2, -a^2, -a^2) . \tag{11}$$

Henceforth the subscripts will be dropped and the particular time coordinate is to be inferred from the context. In this coordinate system the Ricci scalar is found to be

$$R = \sqrt{-g}\,R = -6a^5\ddot{a} - 24a^4\dot{a}^2. \tag{12}$$

Observing that

$$-6a^5\ddot{a} = -6\frac{\mathrm{d}}{\mathrm{d}t}\left(a^5\dot{a}\right) + 30a^4\dot{a}^2, \tag{13}$$

and since the scale factor is held constant on the boundary during extremization, one can just as well vary

$$\sqrt{-g}\,R \rightarrow 6a^4\dot{a}^2. \tag{14}$$

Defining $L = \int \mathrm{d}^4x\,\mathcal{L}$, Eq. (10) becomes

$$\mathcal{L} = \mathcal{L}_{gvm} + \mathcal{L}_{EM} \tag{15}$$

where

$$\mathcal{L}_{gvm} = -\frac{3a^4\dot{a}^2}{8\pi G} - \rho_{v0} - \frac{\rho_{m0}}{a^3} \tag{16}$$

is the gravitational, vacuum, and matter contribution to the action density, and

$$\mathcal{L}_{EM} = -\frac{1}{4}F_{ab}F_{cd}g^{ac}g^{bd} = \frac{1}{2}\left(a^4\mathbf{E}^2 - \frac{\mathbf{B}^2}{a^4}\right) \tag{17}$$

is the electromagnetic contribution to the action density. **B** and **E** are the six degrees of freedom in the anti-symmetric covariant Faraday tensor $F_{ab}$. These fields hide their own dependence on the scale factor which is specific to the context here in which they are components of a radiation field defined in the constant volume system. One could simply infer their dependence by borrowing and transforming from the more familiar form of the Cosmic Microwave Background (CMB) EM energy density in the RW coordinate system. A method more consistent with the Hamiltonian approach of this paper however (albeit principally focused on the scale factor) requires identification of the independent degrees of freedom and their conjugates. The details are given in Appendix B.

*2.3. Hamiltonian*

The first integral of the Euler equations for the total system can be obtained directly from the total Hamiltonian density

$$\rho_{total} \equiv \mathcal{H}_{gvm} + \mathcal{H}_{EM} \tag{18}$$

where

$$\mathcal{H}_{gvm} = \dot{a}\frac{\partial \mathcal{L}_{gvm}}{\partial \dot{a}} - \mathcal{L}_{gvm} = -\frac{3}{8\pi G}a^4\dot{a}^2 + \rho_{v0} + \frac{\rho_{m0}}{a^3} \tag{19}$$

and $\mathcal{H}_{EM}$ is computed in Appendix B. In total therefore

$$\mathcal{H}_{total} = -\frac{3}{8\pi G}a^4 \dot{a}^2 + \rho_{v0} + \frac{\rho_{m0}}{a^3} + \frac{\rho_{r0}}{a^4}. \tag{20}$$

The total energy density $\mathcal{H}_{total}$ is a constant of the motion (independent of the time-dependent scale factor), and appears as an offset to the vacuum energy density $\rho_{v0}$ as a consequence of the particular (constant volume) coordinate system chosen here. Since neither are known a priori, the two can be combined into $\rho_{v0}^*$ say,

$$\rho_{v0}^* \equiv \rho_{v0} - \mathcal{H}_{total}, \tag{21}$$

leaving the number of unknowns unchanged. With this, one finally obtains the Friedmann equation in the constant volume coordinate system

$$-\frac{3}{8\pi G}a^4 \dot{a}^2 + \rho_{v0}^* + \frac{\rho_{m0}}{a^3} + \frac{\rho_{r0}}{a^4} = 0. \tag{22}$$

*2.4. Conservation of energy*

It has been shown that the system in Minkowski space-time

$$L = L_g + L_v + L_m + L_{EM} = -\int d^4x \left\{ \frac{3}{8\pi G}a^4 \dot{a}^2 + \rho_{v0} + \frac{\rho_{m0}}{a^3} + \frac{1}{2}\left(a^4 \mathbf{E}^2 - \frac{\mathbf{B}^2}{a^4}\right) \right\} \tag{23}$$

with dynamical variables $a(t)$ and $\{\phi(t,\mathbf{x}), \mathbf{A}(t,\mathbf{x})\}$ evolves as Eq. (22). Accordingly each term can now be interpreted as an energy density in a Minkowski background. Eq. (22) is just the (first) Friedmann equation expressed in the RW coordinate system Eq. (1) transformed by Eq. (11), so the results are entirely consistent with GR. However, unlike the traditional interpretation of GR, Hamilton's method permits the interpretation of this equation as a statement of energy conservation. This is true independent of the particular coordinate system, though the functional form of the individual terms (matter, vacuum, …) is not.

*2.5. Cosmological expansion*

States of expansion and contraction are indicated by the sign of $da/dt$ in the coordinate system under consideration. If attention is restricted to transformations of the form $d\bar{t} = dt_{RW} a^\beta$ with $a \geq 0$ then 'expansion' and any stationary points have a coordinate independent meaning. Specifically, expansion as commonly understood in the RW coordinate system preserves its meaning in the constant volume system. We should point out however that 'spatial expansion' is not a generally coordinate independent notion – as emphasized recently by Peacock [5]. (See also Barnes et al [6] for related analysis and an interesting discussion.) In the intrinsic time coordinate system Eq. (31) for example, there can be no contracting phase of space, regardless of its presence or absence in the RW coordinate system.

*2.6. Interaction between the fields*

In Eq. (22) all the terms apart from the gravitational field energy are positive and collectively decrease with expansion. It follows that the *magnitude* of the energy in the cosmological gravitational field falls with expansion (the asymptotic behavior is $\dot{a}^2 \propto a^{-4} \Rightarrow a \sim t^{1/3}$ in the constant volume system). That is, the (signed) energy in the gravitational field increases. This increase comes at the expense of matter and radiation. In the constant volume system therefore, *matter and EM fields do work against the gravitational field* during the expansion. Specifically, the price paid by matter is the fall in coordinate energy density by a factor of $1/a^3$, and the price paid by radiation is decay by a factor $1/a^4$; both of these fields *decay* during the expansion. Equivalently, in this coordinate system the expansion might be explained as driven by the matter and radiation fields, the vacuum being neutral in this respect.[2]

Since the end of the radiation-dominated era matter and radiation have been approximately decoupled on a cosmological scale; the mean free path of a photon is of the order of the Hubble radius. From the Hamiltonian perspective however, the very existence of a Friedmann equation implies coupling between the various components. In particular, EM radiation remains coupled to the gravitational field in the present era. In this context one could say that the observation of cosmological red-shift is due to the difference in the way that matter and EM radiation couples to the cosmological gravitational field. Further, in this picture matter and EM radiation remain coupled in the post-recombination era – albeit indirectly – via their participation in the Friedmann equation, now regarded as a constant energy constraint.

*2.7. Gravitational field energy*

In the Hamiltonian framework it follows from the above that the first term in Eq. (22) is the energy associated with the gravitational field. This is to be contrasted with the traditional interpretation of GR where it is difficult to make definite statements about gravitational energy outside the linearized theory. It is of no (physical) consequence that the gravitational energy is coordinate dependent. What matters is that the interpretation of the associated Friedmann equation is self-consistent within each coordinate system; each coordinate system has its own internally self-consistent explanation for energy exchange between the terms in the corresponding Friedmann equation.

---

[2] The system Eq. (22) is closed and time symmetric, so any arrow of time must be attributed to initial conditions and possibly the present phase. Even so, since it has just a single dynamical coordinate, the system cannot describe changes in entropy. Hence there is no substance to the implication of a logically causal flow here.

One is not at liberty to change the sign of all the terms in Eq. (22) in order to facilitate a more familiar interpretation wherein the gravitational field has a positive energy and the other terms represent negative potential energies. It is true that the action is extremized but not minimized by Euler-Lagrange variation so that an overall change in sign can have no impact on the dynamics. But the re-interpretation of mass and EM fields as having negative energy is dismissed as being otherwise unacceptable. The Hamilton Eq. (22) gives that the energy in the cosmological gravitational field is negative. The sign cannot be changed by any transformation of the time coordinate. A special case is what might be called the 'intrinsic' time coordinate system given by $dt = dt_{cv} \dot{a}$ which time is just the scale factor itself – see section 6 below.

With reference to Eq. (22), the magnitude of the gravitational field energy is a monotonically decreasing function of the scale factor. That is, the gravitational field energy tends to zero from below as the universe expands. The energy for this net increase comes from matter and radiation. The evolution of the various components is shown in figure 1. The present is the origin of the time axis, so that $a(0) = 1$.

*2.8. Cosmological red-shift*

Cosmological red-shift is due to a difference (more accurately the ratio) between the matter and radiation dependencies on scale factor. The Hamiltonian-based explanation of how stellar radiation is red-shifted depends on the coordinate system however. Always, the frequency of stellar radiation, e.g. that of Lyman alpha lines, is proportional to the mass of the electron at the site of generation.

In the constant volume system the matter energy density falls like $\rho_{m0}/a^3$ and therefore the Compton frequency of the electron falls likewise.[3] By contrast, the energy density of EM radiation 'in flight' between source and detector falls like $\rho_{r0}/a^4$. The explanation for cosmological red-shift in the constant volume system therefore is that the photon energy associated with a particular Lyman alpha line is reduced by a factor of $1/a$ at the time of reception compared with the energy at the time of emission.

---

3 Strictly, $\rho_{m0}/a^3$ is the contribution to the Hamiltonian from a zero-pressure completely homogeneous 'fluid' so that, in going to a particle description (of which the fluid must therefore be an approximation), this fall in density does not, by itself, say what are the relative contributions of the decay from particle annihilation and mass reduction. The former is excluded, however, under the assumption of decoupling in the post-recombination era, leaving only mass reduction to explain the $\rho_{m0}/a^3$ dependence.

Clearly, as an observed phenomenon, the phrase 'cosmological red-shift' has a corresponding coordinate independent theoretical meaning only if understood to apply to the ratio of the decay rates of matter and radiation with expansion. It does not apply to radiation alone, though this happens to be the case in the particular RW coordinate system. All this is fairly obvious once drawn to attention. One wonders however to what extent the pictures in, for example, Misner, Thorne and Wheeler [7] have given rise to a tendency to regard an expansion of the wavelength of light as a physical - i.e. coordinate independent – description of cosmological red-shift.

It is likely the statements by Harrison [8], Peebles [2] and Baryshev [3] on energy non-conservation are based on the perspective of the Friedmann equation in RW coordinates, though of course with no energy attributed to the gravitational field, in contrast with Hamilton's method as presented here. Since Hamilton's method guarantees conservation of energy it resolves whatever mystery they may have been as to where the energy of EM radiation goes as the universe expands. And though the explanation is different in each coordinate system, it is common to all that Hamilton's method permits the interpretation of the Friedmann equation as a statement of global energy conservation.

*2.9. Vacuum energy*

It is peculiar to the constant-volume coordinate system that the vacuum energy density is unchanged by the expansion.[4] Provided one regards the Cosmological term in the Einstein equation on the same footing as a constant of integration, the constant-volume system has the advantage that the first integral of the equation of motion (the Hamiltonian) does not introduce an arbitrary constant that is not already present in the Einstein equation; the Friedmann equation is reproduced exactly without input of additional assumptions (see section 8 for more details). The case for a Hamiltonian based on the constant-volume coordinate system is perhaps a little stronger therefore than for the conformal and Robertson-Walker systems.[5] Even so, a stronger argument can be made in favor of the Harrison-Newton system discussed in section 5.

---

[4] It can be inferred that the de Sitter gravitational energy density $-3H_0^2/8\pi G$ is a constant in this coordinate system.

[5] Here we allude to the fact that $\Lambda g_{ab}$ is traditionally written on the left hand side of Einstein's equation rather than associated with the stress energy tensor on the right, and therefore represents an arbitrary

**3. Robertson-walker coordinates**

*3.1. General properties*

If the Hamiltonian method as described above is carried out ab initio in the RW coordinate system Eq. (1) the resulting Hamiltonian is found to be

$$-\frac{3}{8\pi G}a\dot{a}^2 + \rho_{v0}a^3 + \rho_{m0}^* + \frac{\rho_{r0}}{a} = 0. \tag{24}$$

In this system the constant of the motion is absorbed into the mass energy density. The time coordinate is now that of Eq. (1) and is related to the constant volume time of Eq. (22) by Eq. (11). By itself however, Eq. (11) is not quite enough to derive Eq. (24) from Eq. (22); here we are concerned to preserve the status of the individual terms in Eq. (24) as energy densities, which requires execution of the Hamiltonian formalism in order to determine an overall common factor (as a power of the scale factor) which would be otherwise undecidable. Eq. (24) is to be contrasted then with the perhaps more familiar form of the Friedmann equation in these coordinates,

$$\frac{3}{8\pi G}\dot{a}^2 = \frac{\rho_{r0}}{a^2} + \frac{\rho_{m0}}{a} + \rho_{v0}a^2. \tag{25}$$

Though mathematically equivalent to Eq. (24), the individual terms in Eq. (25) do not qualify as energy densities within the Hamiltonian framework.

*3.2. Energy balance*

The evolution of the various components of Eq. (24) is shown in figure 2. One sees that the gravitational field energy is not in general a monotonic function of the scale factor. The magnitude of the gravitational field energy is a monotonically increasing function of the scale factor post recombination however, i.e. when $a > \rho_{r0}/\rho_{m0}^*$. In that era, the gravitational field energy becomes increasingly negative as the vacuum energy increases.

Again confining attention to Eq. (24), cosmological red-shift in the RW system is explained as due to a fixed electron Compton frequency (keeping the frequency of the Lyman alpha lines constant during expansion) and a radiation energy density that falls as $1/a$. The RW coordinate system therefore has the advantage that it preserves the invariance of the rest mass familiar from special relativity (i.e. a Lorentz scalar). The disadvantages are that the Maxwell equations for the EM fields are not unchanged and the vacuum modes loose energy.

---

degree of freedom in the governing equations (for *g*) rather than in the source terms or boundary conditions established by matter and radiation.

Since the vacuum energy density is fixed in the constant volume system it does not give rise to urgent concerns about energy conservation. But in the RW system the vacuum energy density scales as $a^3$, (not $a^2$ - see above regarding Eq. (24) versus Eq. (25)) so that the energy of the vacuum is not constant but grows exponentially when $a$ is large. This has caused some to claim that vacuum energy appears from nowhere during the expansion, and to conclude that energy is not conserved at the Cosmological scale. Again, as in the case of Cosmological red-shift, this claim seems to be depend on an interpretation of the individual terms in the Friedmann equation in RW coordinates as signifying energy densities, yet without attributing any energy to the gravitational field. By contrast, and independent of the coordinate system, here we take the position that the Friedmann equation is a statement of energy conservation when viewed from the perspective of Hamilton's method. Consequently there can be no increase in vacuum energy without energy loss elsewhere. Specifically, and with reference to Eq. (24), in the RW system the energy of the vacuum increases at the expense of gravitation and radiation, the latter contribution tending to zero at later times. In the RW system then, one might say that the gravitational field is unstable and spontaneously decays, yielding vacuum energy.

## 4. Conformal coordinates

### 4.1. General properties

The line element in the conformal system is

$$\mathrm{d}s^2 = a^2(t)\left(\mathrm{d}t^2 - \mathrm{d}\mathbf{x}^2\right) \tag{26}$$

for which the corresponding Hamilton-Friedmann equation is

$$-\frac{3}{8\pi G}\dot{a}^2 + \rho_{v0}a^4 + \rho_{m0}a + \rho_{r0}^* = 0. \tag{27}$$

The Maxwell equations are particularly simple in conformal coordinates. The coordinate speed of light is the proper speed of light and is constant throughout the expansion. The system is therefore preferred when analyzing EM interactions, even though the behavior of mass and Compton frequencies are more usually discussed in the framework of the RW coordinate system.

### 4.2. Energy balance

The evolution of the various components is shown in figure 3. From the structure of Eq. (27) one sees that the gravitational field energy is a monotonically decreasing function of expansion at all times. The decaying gravitational field supplies energy to the matter and vacuum fields, the latter dominating at later stages.

In this system the EM radiation modes do not loose energy with propagation, and the Cosmological radiation energy density (primarily that of the CMB) remains unchanged in conformal time. By contrast the classical electron mass increases in proportion to $a(t)$. This result remains true when matter is quantized under an adiabatic approximation involving the scale factor (see Appendix A). In that case, the Compton frequency of the electron increases in proportion to the scale factor. Cosmological red-shift therefore is due not to loss of EM energy but due to increase in the rest mass of inert matter.

Vacuum energy increases uniformly with expansion, the energy for which comes entirely from decay of the gravitational field.

## 5. Newton-harrison system

### 5.1. General properties

If the Hamiltonian method is applied with coordinate choice

$$t_{NH} = \int dt_{RW}\, a(t_{RW}) \Rightarrow \{g_{\mu\nu}\} = diag\left(1/a^2, -a^2, -a^2, -a^2\right) \tag{28}$$

then one obtains

$$-\frac{3}{8\pi G}a^2\left(\frac{da}{dt_{NH}}\right)^2 + \rho_{v0}a^2 + \frac{\rho_{m0}}{a} + \frac{\rho_{r0}}{a^2} + \rho_K = 0. \tag{29}$$

In this coordinate system the constant of the motion $\rho_K$ plays the same role as the curvature term associated in the Robertson-Walker system with the magnitude of the coefficient of *K* in the Friedmann equation. We have called this the Newton-Harrison system because the matter term scales as in the Newtonian Cosmology presented by Harrison [1,9] and others [10-13]. Actually Harrison's Friedmann equation is Eq. (24) - the Robertson-Walker version divided throughout by $a$. The important quality from our point of view though is that each of the terms in Eq. (29) represents an energy density only in the coordinate system Eq. (28).

In this coordinate system the constant of integration cannot be absorbed into any of the matter, vacuum, or radiation contributions. In principle its value is determinable directly from observation of the history of the scale factor (on the backwards light cone from our point of observation). Here such observations are to be interpreted through the lens of a scale-factor field in Minkowski spacetime, with corresponding implications for calculations of luminosity versus distance. Even so, the determination that $\rho_K = 0$ coincides with the same conclusion inferred using the full machinery of GR admitting different cosmological geometries. That is, in both systems

$$-\frac{3}{8\pi G}a^2\left(\frac{da}{dt_{NH}}\right)^2 + \rho_{v0}a^2 + \frac{\rho_{m0}}{a} + \frac{\rho_{r0}}{a^2} = 0 \tag{30}$$

are observationally equivalent predictions, even if departures have different observational consequences. The determination – through the lens of GR – that $K = 0$ in the RW formulation is consistent therefore with the observation that $\rho_K$ - the constant of the motion in the Hamiltonian method – is zero. The fact both $K$ and $\rho_K$ are directly determinable and both are found to be zero argues for adoption of the Newton-Harrison system over the other coordinate systems.

*5.2. Energy balance*

The evolution of the various components is shown in figure 4. Note that just as in the Robertson-Walker system the magnitude of the gravitational energy falls rapidly after the big bang, reaching a minimum and increasing relatively slowly thereafter. All other components are monotonic with expansion. The Compton frequency of matter falls with expansion - assuming as usual that the proper number density is constant. The energy of radiation falls faster with expansion than in the RW system. As is true of all the systems, cosmological red-shift is explained by the fact that radiation energy density falls faster than the matter energy density.

## 6. Intrinsic time

We mention briefly that the time coordinate defined by $dt = da = dt_{RW} da / dt_{RW}$ is a special case. It has been called the 'intrinsic time' by Blythe and Isham [14] and is briefly discussed by Zeh [15] (section 5.3 of that work). From Eq. (22) and Eq. (11) the line element in this system is found to be

$$ds^2 = dt^2 \left(\frac{3}{8\pi G}\right) \frac{t^2}{\rho_{v0} t^4 + \rho_{m0} t + \rho_{r0}} - t^2 d\mathbf{x}^2. \tag{31}$$

In this case there is no Hamiltonian because there is no dynamical coordinate to be varied - the Einstein equations are satisfied identically at all times. In particular, in place of the Friedmann equation, the $G^0{}_0$ component is

$$\frac{1}{8\pi G} G^0{}_0 = \rho_{v0} + \rho_{m0}/t^3 + \rho_{r0}/t^4 = T^0{}_0. \tag{32}$$

## 7. Quantum cosmology

A Hamiltonian formulation of cosmology offers the possibility of an easy transition to a quantum theory. A full analysis is beyond the scope of this article; here we offer a few observations leaving the details for a possible future article.

The literature on quantization of the Friedmann equation dates back to de Witt [16] and Misner [17]. A review of the early efforts can be found in the book by Ryan [18]. An outcome of those analyses was the 'Hamiltonian constraint' that the total energy is zero and therefore that the corresponding Schrödinger equation should be $\hat{H}\psi = 0$. The corresponding loss of a role for a time coordinate is the subject of discussion in those and subsequent works, including for example [14,19]. A more recent review of some of these efforts has been given by Zeh [15].

In the formally Minkowski 'field theory' presented here (i.e. with $a(t)$ as the sole dynamical variable in flat spacetime) the Hamiltonian constraint applies only in the state of an empty vacuum (i.e. with no energy and no Cosmological constant). And in particular, in the cosmological case therefore one is free to construct a full (time dependent) Schrödinger equation. In the preferred choice of the Newton-Harrison coordinate system a non-zero total energy corresponds to a departure from flatness, at least in so far as it appears in the Friedmann equation (though not in the luminosity versus distance relations based upon a Robertson-Walker $K = -1$ geometry). Even after fixing the time coordinate there remains however an ambiguity of operator ordering in the canonical quantization of Eq. (29). This could be resolved if one first defines a dynamical coordinate $x(t) = a^2(t)$ to give

$$\mathcal{H}_{total} = -\frac{3}{32\pi G}\left(\frac{dx}{dt_{NH}}\right)^2 + \rho_{v0}x + \frac{\rho_{m0}}{\sqrt{x}} + \frac{\rho_{r0}}{x}. \tag{33}$$

Quantization is now unambiguous,

$$\frac{3\hbar^2}{32\pi G}\frac{\partial^2 \psi}{\partial x^2} + \left(\rho_{v0}x + \frac{\rho_{m0}}{\sqrt{x}} + \frac{\rho_{r0}}{x}\right)\psi = -i\hbar\frac{\partial \psi}{\partial t}, \tag{34}$$

though it is unsatisfactory that the ambiguity is circumvented by an arbitrary substitution rather than confronted more directly and resolved on physical or mathematical grounds. It remains to see if there is any contact between Eq. (34) and the observational facts.

**8. Comparison between Hamiltonian and more general gravitational stress-energy methods**

*8.1. Einstein Tensor as Stress-Energy Tensor*

Since the method of obtaining the stress-energy from a Lagrangian density $\Delta$

$$\frac{1}{2}\sqrt{-g}T_{ab} = \left(\frac{\partial}{\partial g^{ab}} - \partial_c\frac{\partial}{\partial(\partial_c g^{ab})}\right)\sqrt{-g}\Delta \tag{35}$$

is the same as that by which the Hilbert action Eq. (3) is extremized by variation of the individual of components $g^{ab}$ so as to give the Einstein tensor $G_{ab}$ (see for example [20])

$$\frac{1}{16\pi G}\left(\frac{\partial}{\partial g^{ab}} - \partial_c\frac{\partial}{\partial(\partial_c g^{ab})}\right)\sqrt{-g}R = -\frac{1}{16\pi G}\sqrt{-g}G_{ab}; \quad G_{ab} \equiv R_{ab} - \frac{1}{2}g_{ab}R, \tag{36}$$

it follows that Einstein's equations may be interpreted as setting to zero the total stress-energy, the gravitational component of which is

$$T_{ab(g)} = -G_{ab}/8\pi G.  \qquad (37)$$

No new physics flows from this association. $T_{ab(g)}$ vanishes where there are neither matter nor radiation fields, and so does not have the desired characteristics of a tensor to describe the effects of gravitational radiation for example.[6] By contrast, since neither the Cosmological matter nor radiation fields vanish anywhere, $T_{ab(g)}$ is well-suited to give an intuitive account of the gravitational stress-energy in cosmology. And unlike the pseudo-tensor usually employed for gravitational radiation, the eigenvalues of $T^a{}_{b(g)}$ have a coordinate-independent meaning. In particular, $T^0{}_{0(g)} \equiv \rho_g$ is a coordinate energy density that cannot be transformed away. In the approach here, after making the coordinate choice (constant volume, Robertson Walker, conformal, ...), each of the quantities $\rho_g, \rho_r, \rho_v, \rho_m$ is then interpreted as a density in Minkowski spacetime expressed in Lorentz-Minkowski coordinates. From this association it follows that the Hamiltonian procedure and projection onto Minkowski spacetime can be regarded as the specialization of a method to extract a full Lorentz-Minkowski stress energy tensor wherein the metric is decomposed as

$$g_{ab} = \eta_{ac} f^c{}_b. \qquad (38)$$

After making this substitution in the action, $\eta_{ac}$ would be interpreted as the metric of a Lorentz-Minkowski 'background', $f^c{}_b$ would be interpreted as a Lorentz tensor field, and Einstein's equations would follow from variation of the $f^c{}_b$. The decomposition (38) fixes the relationship between the metric and the background only once the metric is expressed in a particular set of coordinates – it is not unique. This approach has only a superficial resemblance to the decomposition of the metric into tetrads $g_{ab} = e_a^{(p)} \eta_{pq} e_b^{(p)}$. One difference is that Eq. (38) establishes a one-to-one relationship between the tensor field $f^c{}_b$ and the metric tensor, from which it can be inferred that Eq. (38) cannot be used to extend general coordinate invariance to spinor fields.

---

[6] The association Eq. (37) is discussed in Landau and Lifshitz in a footnote in section 96 [20]. Therein, the authors warn against employing Eq. (37), presumably because they are concerned exclusively with gravitational radiation, pointing out that this approach implies zero stress-energy in vacuum in that case.

To be more concrete, after having made the substitution (38), one then presumes a Minkowski spacetime with metric $\{\eta_{ab}\} = diag(1,-1,-1,-1)$ with dynamical variable $a(t)$ one expects to be able to write the Lagrangian density $\mathsf{L} = \Lambda\sqrt{-\eta}$ with

$$\Lambda = \eta^{00}\dot{a}^2(t)K(a(t)) - V(a(t)) \tag{39}$$

for some (arbitrary) pair of functions $K, V$. Eq. (35) then gives that the energy density can be found from

$$\frac{1}{2}\sqrt{-\eta}T_{00} = \frac{\partial}{\partial \eta^{00}}\left(\sqrt{-\eta}\Lambda\right) \tag{40}$$

where

$$\frac{\partial\sqrt{-\eta}}{\partial \eta^{00}} = \frac{d\eta_{00}}{d\eta^{00}}\frac{\partial\sqrt{-\eta}}{\partial \eta_{00}} = -\frac{1}{2} \tag{41}$$

and so

$$\frac{1}{2}T_{00} = -\frac{1}{2}\left(\dot{a}^2 K(a) - V(a)\right) + \dot{a}^2 K(a) \Rightarrow T_{00} = T_0^0 = \dot{a}^2 K(a) + V(a). \tag{42}$$

Following the discussion above, it follows that

$$\dot{a}^2 K(a) + V(a) = 0 \tag{43}$$

is the Friedmann equation in the coordinate system in which the decomposition Eq. (38) was implemented. By contrast, the total Hamiltonian density due to Eq. (39) is

$$\mathcal{H} = \dot{a}\frac{\partial\Lambda}{\partial \dot{a}} - \Lambda = \dot{a}^2 K(a) + V(a) \tag{44}$$

and therefore

$$\dot{a}^2 K(a) + V(a) = \text{constant} \tag{45}$$

to be compared with Eq. (43). Hamilton's method therefore introduces a constant that appears to be absent from the Einstein theory. However from the discussion in section 5 it follows that the two approaches can be brought into a one-to-one correspondence in the special case of the Newton-Harrison coordinate system, and perhaps also - depending on one's interpretation of the Cosmological constant – in the constant volume system.

*8.2. Relationship to the methods of Babak, Grischuk, Logunov, and related efforts*

Recently Babak and Grischuk [21] have published a method based upon a split of the metric tensor into a fixed background with metric $\gamma^{ab}$ and a field $h^{ab}$ on that background according to the prescription

$$\sqrt{-g}\,g^{ab} = \sqrt{-\gamma}\left(\gamma^{ab} + h^{ab}\right) \tag{46}$$

where the metric tensor $g^{ab}$ satisfies the Einstein equations as usual. The background $\gamma^{ab}$ is stipulated to be Minkowski though expressed in any coordinate system. Writing Eq. (38) as

$$g^{ab} = \left(f^{-1}\right)^a{}_c \eta^{cb} \tag{47}$$

and combining with Eq. (46) (with $\gamma^{ab} = \eta^{ab}$) gives the $h^{ab}$ field in terms of the $f^c{}_b$ employed here:

$$\sqrt{-\eta}\left(\eta^{ab} + h^{ab}\right) = \sqrt{-g}\left(f^{-1}\right)^a{}_c \eta^{cb} \Rightarrow h^{ab} = \left(\frac{\sqrt{-g}}{\sqrt{-\eta}}\left(f^{-1}\right)^a{}_c - \delta^a{}_c\right)\eta^{cb}. \tag{48}$$

The main result of Babak and Grischuk is extraction of a gravitational stress energy tensor $T^{ab}(h;\gamma)$ from the total action (with the replacement Eq. (46)) by differentiation with respect to the background $\gamma^{ab}$ in the manner of Eq. (38). Because the background is not in a one-to one correspondence with the metric $g^{ab}$, this stress-energy tensor cannot be the Einstein tensor as it is in the generalization of the Hamiltonian method presented here. Broadly, this approach is an attempt to split the Einstein tensor into two pieces: a covariant generalization of the D'Alembertian of the metric, and the remainder. That work is the object of a recent critique [22]. The problem seems to be that because $g^{ab}$ is already underdetermined by a general coordinate transformation, its expression in different coordinates gives rise to different $h^{ab}$ through Eq. (46) and that different members of the allowable family of $h^{ab}$ result in different stress-energy tensors $T^{ab}(h;\gamma)$ derived from variation of the background $\gamma^{ab}$; these $T^{ab}(h;\gamma)$ are not themselves related by a general coordinate transformation.

The method of Babak and Grischuk is very closely related to the work of Logunov [23-25]. The main difference seems to be that Logunov seeks to remove the ambiguity in Eq. (46) with the additional constraint that the metric first be cast in harmonic coordinates, which choice is mandated by a non-zero graviton mass. The harmonic coordinate condition appears in a role analogous to the Lorenz condition in EM, mandated there by a non-zero photon mass.

So far the Logunov method seems to have withstood scrutiny, and has recently received support from Visser [26]. The work of Yilmaz [27,28] appears to be similarly motivated if different in execution, though has been the object of recent critical scrutiny [29]. The technique of splitting the Einstein tensor into a D'Alembertian of the metric plus some non-linear remainder to be designated as the stress-energy of the gravitational field in a flat space background dates back to Rosen [30,31] and Papapetrou [32].

## 9. Conclusion

A Hamiltonian presentation of Friedmann-Robertson-Walker cosmology provides an intuitive framework for understanding energy balance during cosmological expansion. A crucial component is the identification of the energy of the cosmological gravitational field which is otherwise missing from more traditional interpretations.

**Appendix A. Influence of cosmological expansion on a massive scalar field**

Here we briefly look at the consequences of a changing scale factor for the scalar field of a single massive quantum particle. At issue is a comparison with the behavior of a classical point particle whilst taking into account that the quantum particle is in general smeared out over space and therefore potentially sensitive to metric coefficients other than $g_{00}$.

Let the free-field action be

$$L_m = -\int d^4x \sqrt{-g}\left(g^{ab}(\partial_a\phi)(\partial_b\phi) + (m^2 + \xi R)\phi^2\right) \tag{A.1}$$

where $c = \hbar = 1$, and where $\xi$ is a dimensionless parameter that controls the coupling to the curvature. Assuming, initially, a general conformal metric - thereby covering all three RW geometries [33] -

$$g_{ab}(x) = \eta_{ab} f^2(x), \tag{A.2}$$

the Ricci scalar curvature is found to be

$$R = 6\eta^{ab}(\partial_a f \partial_b f)/f^2. \tag{A.3}$$

The Euler equations for variation of $\phi$ are therefore

$$\eta^{ab}\partial_a(f^2\partial_b\phi) + (f^2 m^2 + 6\xi\eta^{ab}\partial_a f \partial_b f)f^2\phi = 0. \tag{A.4}$$

The substitution $\phi(x) = \psi(x)/f(x)$ in Eq. (A.4) eliminates the first derivatives to give

$$\partial^2\psi + \left(m^2 f^2 + 6\xi(\partial f)^2 - \frac{1}{f}\partial^2 f\right)\psi = 0 \tag{A.5}$$

where

$$\partial^2 \equiv \eta^{ab}\partial_a\partial_b, \quad (\partial f)^2 \equiv \eta^{ab}\partial_a f \partial_b f. \tag{A.6}$$

Eq. (A.5) can be interpreted as a Klein-Gordan equation in Minkowski space-time with a variable mass. The 4-vector

$$j_a = \phi^*\partial_a\phi - \phi\partial_a\phi^* = (\psi^*\partial_a\psi - \psi\partial_a\psi^*)/f^2 \tag{A.7}$$

has covariant divergence

$$j_a^{;a} = \partial_b(\sqrt{-g}g^{ab}j_a) = \eta^{ab}\partial_b(f^2 j_a) = \eta^{ab}\partial_b(\psi^*\partial_a\psi - \psi\partial_a\psi^*). \tag{A.8}$$

The latter vanishes due to Eq. (A.5), therefore Eq. (A.7) is a candidate for the 4-current. Consistent with the perspective presented elsewhere in this document, $\psi$ is to be regarded as a quantum wavefunction in Minkowski spacetime whose associated current is independently conserved, though coupled to the cosmological gravitational field $f(x)$ as it appears in Eq. (A.5).

In the case of the $K = 0$ geometry in conformal time $f(x) = a(t)$ and Eq. (A.5) reduces to

$$\partial^2 \psi + \left(m^2 a^2 + 6\xi \dot{a}^2 - \ddot{a}/a\right)\psi = 0. \tag{A.9}$$

Using that in conformal coordinates

$$\dot{a} = H(t) \Rightarrow \ddot{a} = \left(1 - q(t)\right) H^2(t) a \tag{A.10}$$

where $H$ is the Hubble parameter and $q$ is the deceleration parameter, then

$$\partial^2 \psi + a^2 \left(m^2 + \left(6\xi + 1 - q\right) H^2\right)\psi = 0. \tag{A.11}$$

One sees from Eq. (A.11) that a sufficient condition for the mass of a particle to be accurately characterized as increasing in proportion to the scale-factor is that $\left(6\xi + 1 - q\right) H^2 \ll m^2$ and the fractional variation of the scale factor over the interval of interest is very small. These conditions are easily met by any reasonable post-recombination wavefunction. In such cases, at any time $t_{era}$ Eq. (A.11) can be interpreted as the traditional Klein-Gordon equation in Minkowski spacetime with constant (though adiabatically increasing) mass $m^* = a(t_{era}) m$.

**Appendix B. Energy density of cosmological EM radiation fields in the constant volume system**

In this appendix the energy density of the cosmological EM field - to be identified principally with the CMB – is computed within the Hamiltonian formalism. We will take it that the independent degrees of freedom are the potentials satisfying

$$\mathbf{B} = \nabla \times \mathbf{A}, \quad \mathbf{E} = -\nabla \phi - \frac{\partial \mathbf{A}}{\partial t}. \tag{B.1}$$

Putting this in Eq. (17) gives that the EM contribution to the action is

$$\mathcal{L}_{EM} = -\frac{1}{2}\left(\frac{(\nabla \times \mathbf{A})^2}{a^4} - a^4 \left(\nabla \phi + \frac{\partial \mathbf{A}}{\partial t}\right)^2\right). \tag{B.2}$$

If we designate $\phi, \mathbf{A}$ as the independent components, then the Hamiltonian density is

$$\begin{aligned}\mathcal{H}_{EM} &= \frac{\partial \phi}{\partial t}\frac{\partial \mathcal{L}_{EM}}{\partial(\partial \phi/\partial t)}+\frac{\partial \mathbf{A}}{\partial t}\cdot\frac{\partial \mathcal{L}_{EM}}{\partial(\partial \mathbf{A}/\partial t)}-\mathcal{L}_{EM} \\ &= a^4\frac{\partial \mathbf{A}}{\partial t}\cdot\left(\nabla\phi+\frac{\partial \mathbf{A}}{\partial t}\right)+\frac{1}{2}\left(\frac{(\nabla\times\mathbf{A})^2}{a^4}-a^4\left(\nabla\phi+\frac{\partial \mathbf{A}}{\partial t}\right)^2\right) \\ &= a^4\left(\mathbf{E}^2+\mathbf{E}\cdot\nabla\phi\right)+\frac{1}{2}\left(\frac{\mathbf{B}^2}{a^4}-a^4\mathbf{E}^2\right) \\ &= \nabla\left(a^4\phi\mathbf{E}\right)+\frac{1}{2}\left(\frac{\mathbf{B}^2}{a^4}+a^4\mathbf{E}^2\right)\end{aligned}$$

(B.3)

The last step follows because we are discussing vacuum fields, so there is no local charge. And since we are assuming a constant spatial energy density, the result will be unchanged if the energy density is replaced by its spatial average. The first term then contributes only a boundary (surface) term which we will assume can be neglected, so that

$$\langle \mathcal{H}_{EM}\rangle = \frac{1}{2}\left\langle\frac{\mathbf{B}^2}{a^4}+a^4\mathbf{E}^2\right\rangle$$

(B.4)

the angle brackets denoting a spatial average.[7]

It remains to determine the dependency of the electric and magnetic fields on the scale factor. This seems to require explicit construction of vacuum solutions to the Maxwell equations in the chosen (unit determinant) coordinate system. Variation of $\phi$ gives

$$-a^4\nabla\left(\nabla\phi+\frac{\partial \mathbf{A}}{\partial t}\right)=0 \Rightarrow \partial^2\phi=0$$

(B.5)

in the Lorenz Gauge, whilst variation of $\mathbf{A}$ gives

$$-\frac{\nabla\times(\nabla\times\mathbf{A})}{a^4}-\frac{\partial}{\partial t}\left(a^4\left(\nabla\phi+\frac{\partial \mathbf{A}}{\partial t}\right)\right)=0.$$

(B.6)

For the magnetic field one therefore has

$$-\frac{\nabla \times \mathbf{B}}{a^4} + \frac{\partial}{\partial t}(a^4 \mathbf{E}) = 0$$

$$\Rightarrow -\nabla \times (\nabla \times \mathbf{B}) + a^4 \frac{\partial}{\partial t}\left(a^4 (\nabla \times \mathbf{E})\right) = 0$$

$$\Rightarrow \nabla^2 \mathbf{B} - a^4 \frac{\partial}{\partial t}\left(a^4 \left(\nabla \times \frac{\partial \mathbf{A}}{\partial t}\right)\right) = 0 \quad \text{(B.7)}$$

$$\Rightarrow \nabla^2 \mathbf{B} - a^4 \frac{\partial}{\partial t}\left(a^4 \frac{\partial \mathbf{B}}{\partial t}\right) = 0$$

the general solution of which can be written

$$\mathbf{B} = \operatorname{Re} \sum_{\mathbf{k}} \mathbf{b_k} \exp\left(i\mathbf{k}.\mathbf{x} - i|\mathbf{k}|\int dt/a^4\right). \quad \text{(B.8)}$$

---

[7] If we were not constrained to use the Hamiltonian approach, this result could have been found more directly by differentiation with respect to the metric of the electromagnetic part of the action [20]. Specifically, one has $T^a{}_b = -F^a{}_c F_b{}^c + \frac{1}{4}F_{cd}F^{cd}\delta^a{}_b = g^{-1}Fg^{-1}F - \frac{1}{4}tr\left(g^{-1}Fg^{-1}F\right)\mathbf{1}$ where $F$ with no indexes is the covariant Faraday tensor. Putting in from Eq. (11) then gives $\mathcal{H}_{EM} \equiv T^0{}_0 = \frac{1}{2}\left(a^4\mathbf{E}^2+\mathbf{B}^2/a^4\right)$ with much less trouble.

And for the electric field one has

$$\frac{-\nabla(\nabla.\mathbf{A}) + \nabla^2 \mathbf{A}}{a^4} - \frac{\partial}{\partial t}\left(a^4\left(\nabla\phi + \frac{\partial \mathbf{A}}{\partial t}\right)\right) = 0$$

$$\Rightarrow -\nabla(\nabla.(\mathbf{E}+\nabla\phi)) + \nabla^2(\mathbf{E}+\nabla\phi) + \frac{\partial}{\partial t}\left(a^4 \frac{\partial}{\partial t}(a^4 \mathbf{E})\right) = 0 \quad \text{(B.9)}$$

$$\Rightarrow \nabla^2 \mathbf{E} - \frac{\partial}{\partial t}\left(a^4 \frac{\partial}{\partial t}(a^4 \mathbf{E})\right) = 0$$

the general solution of which can be written

$$\mathbf{E} = \operatorname{Re} \sum_{\mathbf{k}} \frac{\mathbf{e_k}}{a^4} \exp\left(i\mathbf{k}.\mathbf{x} - i|\mathbf{k}|\int dt/a^4\right). \quad \text{(B.10)}$$

A characteristic of the unit determinant coordinate system therefore is that the Fourier components of the magnetic field are unaffected by the expansion, whereas the components of the electric field fall as $1/a^4$. These differing responses to the expansion cancel the effect of the unequal weighting in Eq. (B.4), so that for fixed $\mathbf{k}$ the magnetic and electric Fourier components contribute equally to the energy:

$$\langle \mathcal{H}_{EM} \rangle = \langle \rho_r(a) \rangle = \frac{1}{2a^4} \sum_{\mathbf{k}} \left(\mathbf{e_k}^2 + \mathbf{b_k}^2\right) = \frac{\rho_{r0}}{a^4} \quad \text{(B.11)}$$

where $\rho_{r0}$ is the EM energy density at the time $a=1$. Note that this result is specific to the unit-determinant coordinate system. It could however have been inferred from the well-known expression for the EM energy density as it appears in the Friedmann equation written, most commonly, in the RW coordinate system.

**References**

[1] Harrison E R 2000 *Cosmology: The Science of the Universe* (Cambridge, UK: Cambridge University Press)
[2] Peebles P J E 1993 *Principles of Physical Cosmology* (Princeton, NJ: Princeton University Press)

[3] Baryshev Y V 2005 Conceptual Problems of the Standard Cosmological Model *AIP Conference Proceedings* **822** eds E J Lerner and J B Almeida pp 23-33
[4] Feynman R P, Morinigo F B, and Wagner W G 1995 *Feynman lectures on Gravitation* (Reading, Massachusetts: Addison-Wesley)
[5] Peacock J A 2008 A diatribe on expanding space *arxiv: 0809.4573*
[6] Barnes L A et al 2006 Joining the Hubble Flow: Implications for Expanding Space *Monthly Notices of the Royal Astronomical Society* **373** 382-90
[7] Misner C W, Thorne K S, and Wheeler J A 1973 *Gravitation* (New York: W. H. Freeman and Co.)
[8] Harrison E R 2000 *Cosmology: the science of the universe* (Cambridge, UK: Cambridge University Press)
[9] Harrison E R 1986 Newton and the infinite universe *Physics Today* **39** 24-30
[10] Layzer D 1954 The significance of Newtonian cosmology *Astronomical Journal* **59** 168
[11] McRea W H and Milne E A 1934 Newtonian universes and the curvature of space *Quart. J. Math.* **5** 73-80
[12] Milne E 1934 A Newtonian expanding universe *Quarterly Journal of Mathematics* **5** 64
[13] McCrea W H 1955 On the significance of Newtonian Cosmology *Astronomical Journal* **60** 271
[14] Blythe W F and Isham C J 1975 Quantization of a Friedmann universe filled with a scalar field *Phys. Rev. D* **11** 768-78
[15] Zeh H D 2007 *The physical basis of the direction of time* (Berlin: Springer)
[16] de Witt B 1967 Quantum Theory of Gravity. I. The Canonical Theory *Phys. Rev.* **160** 1113-48
[17] Misner C W 1969 Quantum Cosmology. 1 *Phys. Rev.* **186** 1319-27
[18] Ryan M 1972 *Hamiltonian Cosmology* (New York: Springer)
[19] Kaup D J and Vitello A P 1974 Solvable quantum cosmological model and the importance of quantizing in a special canonical frame *Phys. Rev. D* **9** 1648-55
[20] Landau L D and Lifshitz E M 2000 *The Classical Theory of Fields* (Oxford, UK: Butterworth-Heinemann)
[21] Babak S V and Grishchuk L P 1999 *Phys. Rev. D* **61** 024038
[22] Butcher L M, Lasenby A, and Hobson M 2008 The physical significance of the Babak-Grishchuk gravitational energy-momentum tensor *Phys. Rev. D* - submitted
[23] Logunov A A 1995 The theory of the classical gravitational field *Physics - Uspekhi* **38** 179-93
[24] Logunov A A 1996 Classical gravitational field theory and Mach principle *Gravity particles and spacetime* eds P Pronin and G Sardanashvily (Singapore: World Scientific) pp 173-205
[25] Logunov A A and Mestvirishvili M 1989 *The relativistic theory of gravitation* (Moscow: Mir Publishers)
[26] Visser M 1998 Mass for the graviton *General Relativity and Gravitation* **30** 1717-28
[27] Yilmaz H 1992 Towards a Field Theory of Gravitation *Nuovo Cimento B* **107** 941-60
[28] Yilmaz H 1976 Physical Foundations of the New Theory of Gravitation *Annals of Physics* **101** 413-32
[29] Ibison M 2006 Cosmological test of the Yilmaz theory of gravity *Class. Quantum Grav.* **23** 577-89
[30] Rosen N 1940 General Relativity and Flat Space. I *Phys. Rev.* **57** 147-50
[31] Rosen N 1940 General Relativity and Flat Space. II *Phys. Rev.* **57** 150-3
[32] Papapetrou A 1948 Einstein's theory of gravitation and flat space *Proc. Roy. Irish Acad.* **52A** 11-23
[33] Ibison M 2007 On the conformal forms of the Robertson-Walker metric *J. Math. Phys.* **48** 122501-1-122501-23

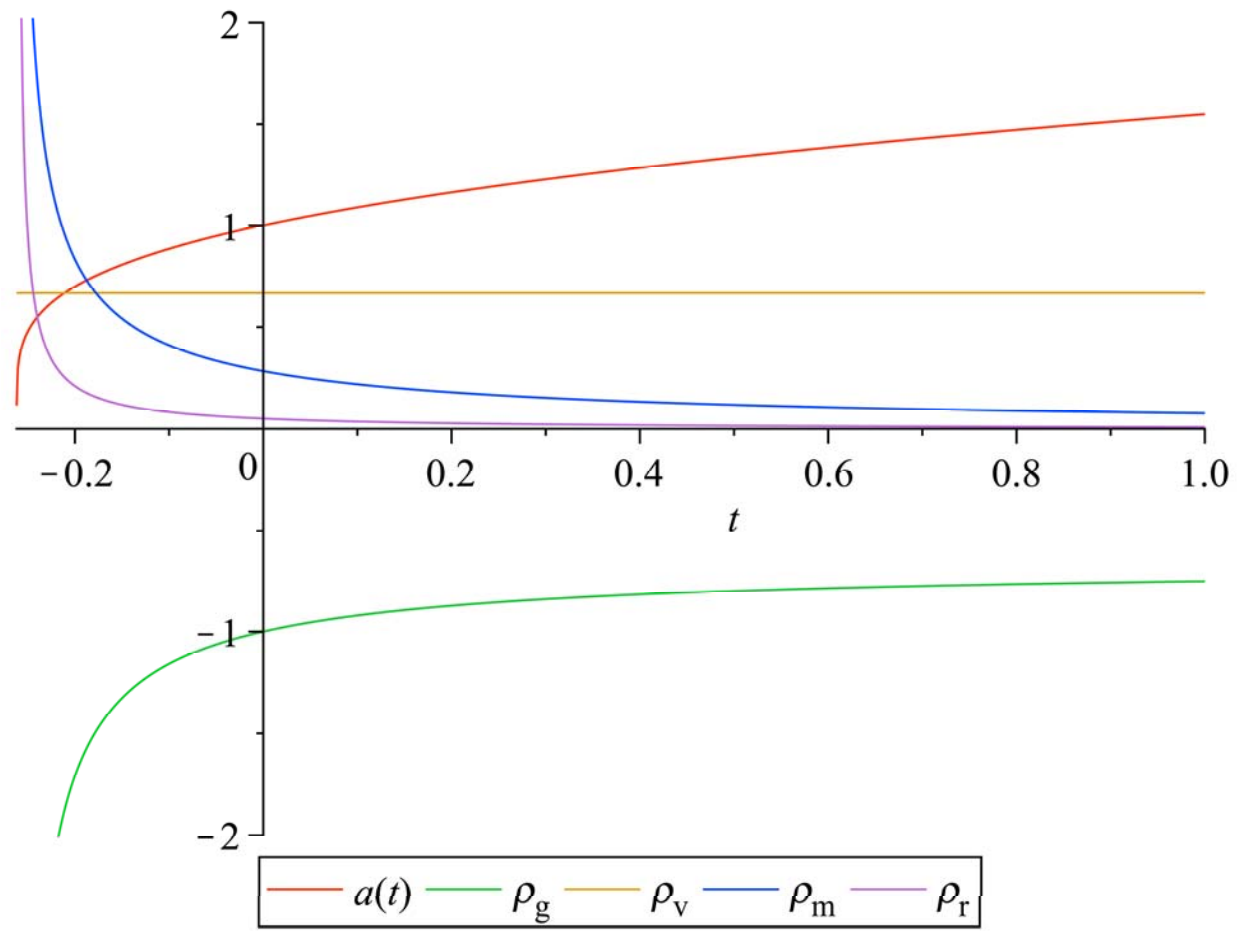

Figure 1: Cosmological evolution in the constant volume system

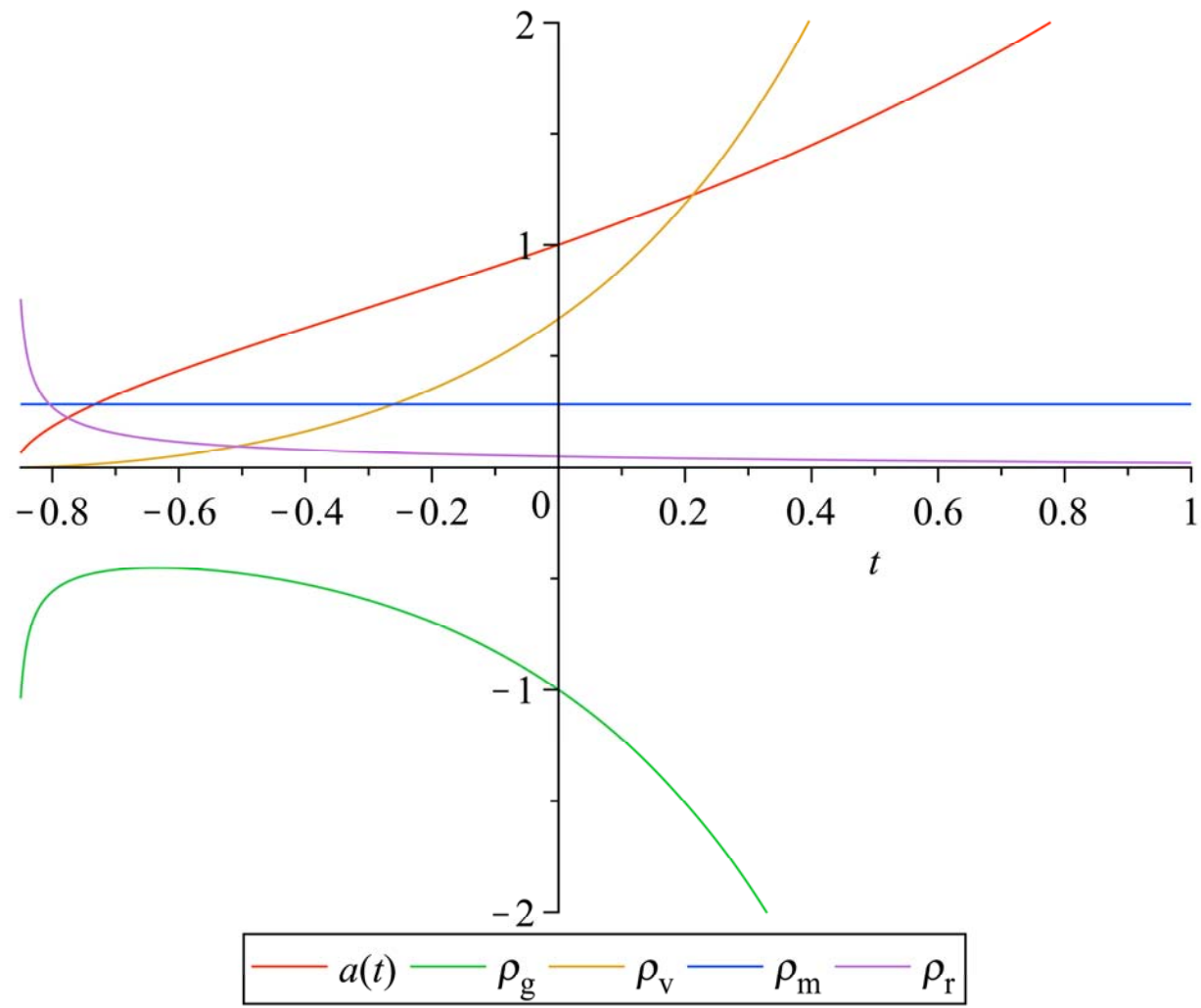

Figure 2: Cosmological evolution in the Robertson-Walker system

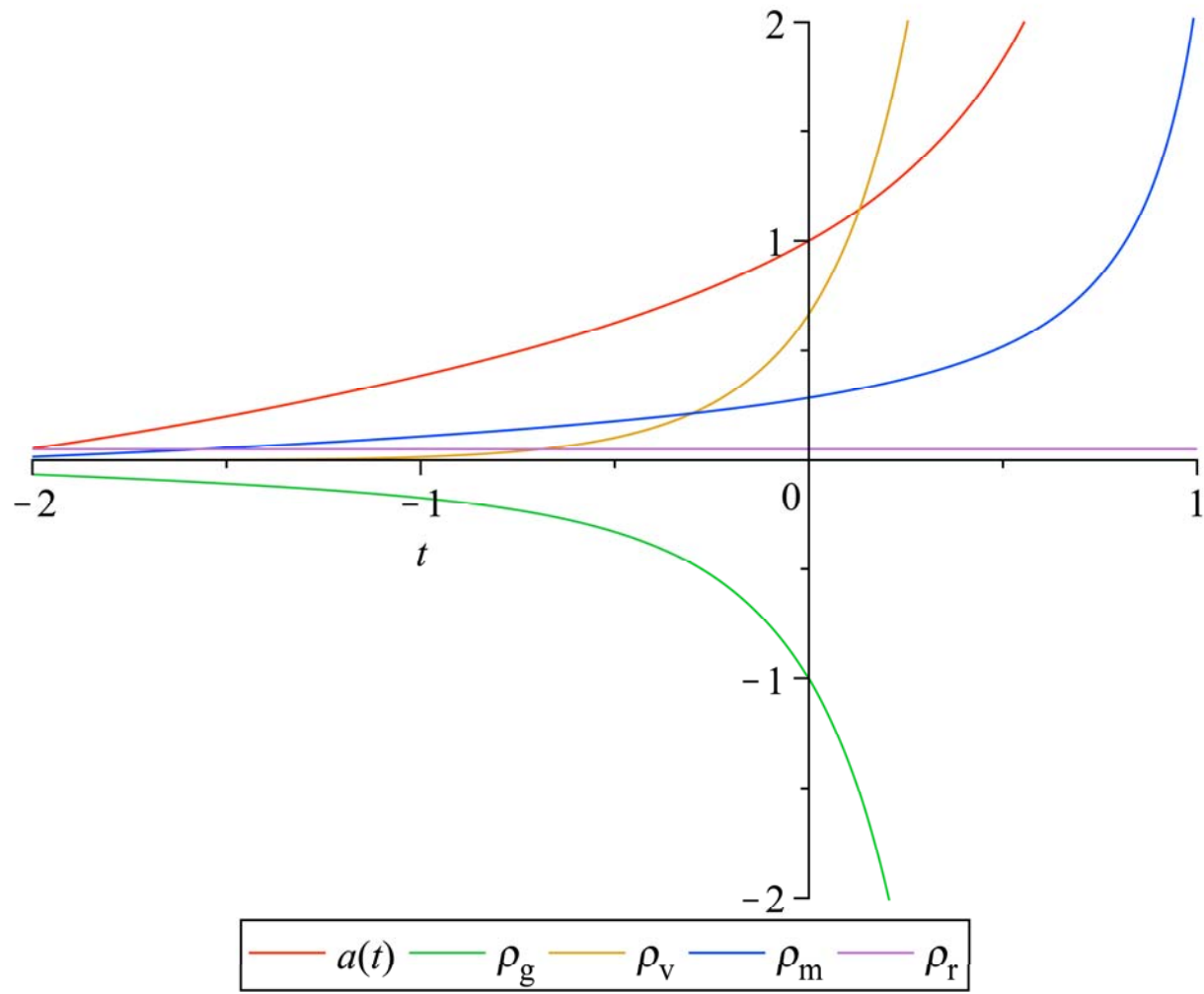

Figure 3: Cosmological evolution in the conformal system

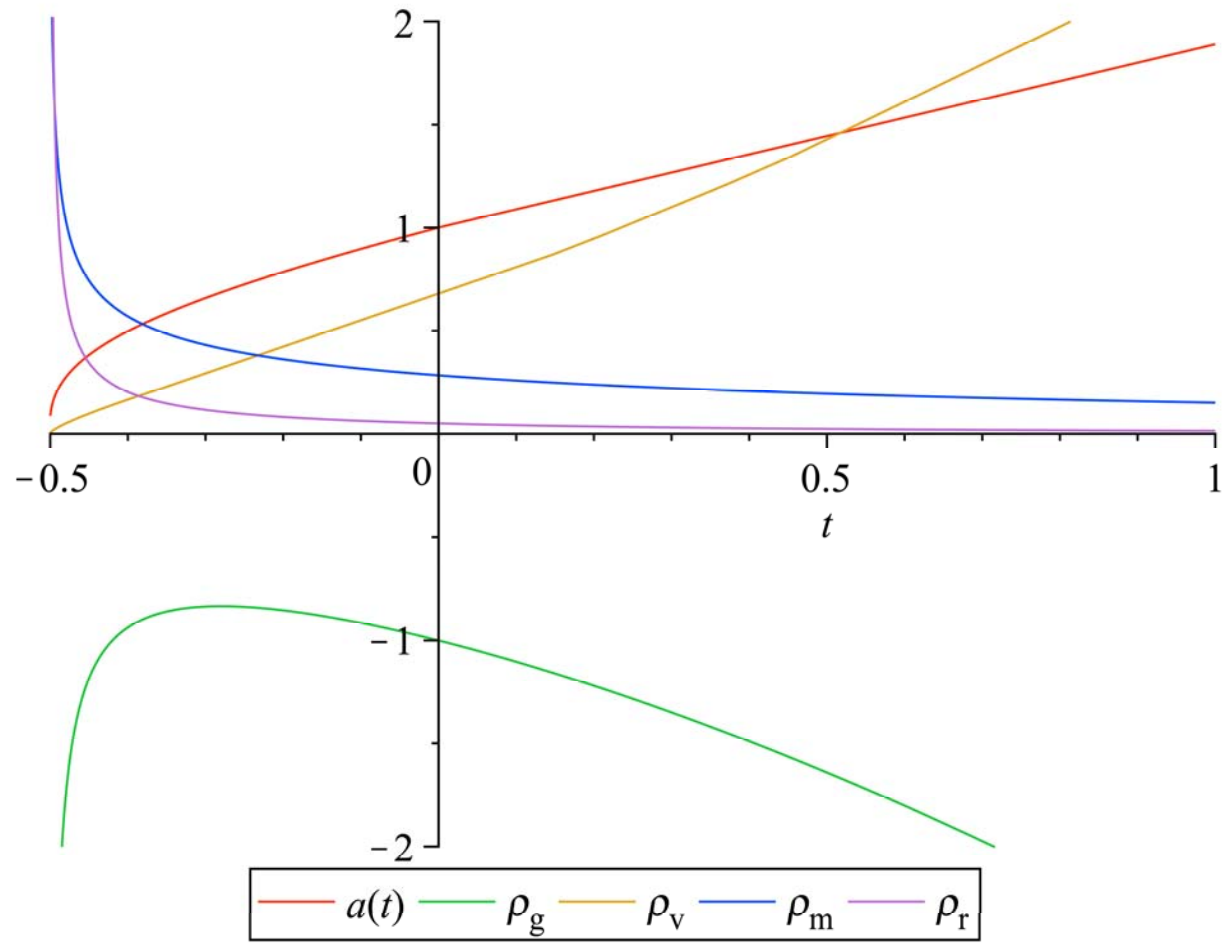

Figure 4: Cosmological evolution in the Newton-Harrison system